\documentclass[letterpaper,twocolumn,10pt]{article}

\usepackage{usenix-2020-09}

\usepackage{amsmath}
\usepackage{microtype}
\usepackage{booktabs}
\usepackage{xspace}
\usepackage{graphicx}
\usepackage{float}
\usepackage{subfigure}
\usepackage{subfloat}
\usepackage{pgfplots}
\usepackage{layouts}
\usepackage{caption}
\usepackage{siunitx}
\usepackage{multirow}
\usepackage{hyperref}
\usepackage{cleveref}
\usepackage{todonotes}
\usepackage{xcolor}
\usepackage{colortbl}
\usepackage{hhline}
\usepackage{enumitem}
\setlist{noitemsep,topsep=0in}
\newcommand{\protocolnamelong}{A Resource Record Fragmentation mechanism\xspace}
\newcommand{\protocolname}{ARRF\xspace}
\newcommand{\lookupname}{example.com\xspace}
\newcommand{\lookuptld}{.com\xspace}

\usetikzlibrary{positioning}

\begin{document}

\date{November 25, 2022}

\title{\Large \bf Post-Quantum Signatures in DNSSEC \\ via Request-Based Fragmentation}

\author{
{\rm Jason Goertzen}\\
University of Waterloo\\
\href{mailto:jgoertze@uwaterloo.ca}{\rm\texttt{jgoertze@uwaterloo.ca}}
\and
{\rm Douglas Stebila}\\
University of Waterloo\\
\href{mailto:dstebila@uwaterloo.ca}{\rm\texttt{dstebila@uwaterloo.ca}}
}

\maketitle

\begin{abstract}
The Domain Name System Security Extensions (DNSSEC) provide authentication of DNS responses using digital signatures.
DNS operates primarily over UDP, which leads to several constraints: notably, DNS packets should be at most 1232 bytes long to avoid problems during transmission. 
Larger DNS responses would either need to be fragmented into several UDP responses or the request would need to be repeated over TCP, neither of which is sufficiently reliable in today's DNS ecosystem.
While RSA or elliptic curve digital signatures are sufficiently small to avoid this problem, even for DNSSEC packets containing both a public key and a signature, this problem is unavoidable when considering the larger sizes of post-quantum schemes.

We propose ARRF, a method of fragmenting DNS resource records at the application layer (rather than the transport layer) that is \emph{request-based}, meaning the initial response contains a truncated fragment and then the requester sends follow-up requests for the remaining fragments. 
Using request-based fragmentation avoids problems identified for several previously proposed---and rejected---application-level DNS fragmentation techniques.
We implement our approach and evaluate its performance in a simulated network when used for the three post-quantum digital signature schemes selected by NIST for standardization (Falcon, Dilithium, and SPHINCS+) at the 128-bit security level.
Our experiments show that our request-based fragmentation approach provides substantially lower resolution times compared to standard DNS over UDP with TCP fallback, for all the tested post-quantum algorithms, and with less data transmitted in the case of both Falcon and Dilithium.
Furthermore, our request-based fragmentation design can be implemented relatively easily: our implementation is in fact a small daemon that can sit in front of a DNS name server or resolver to fragment/reassemble transparently.
As well, our request-based application-level fragmentation over UDP may avoid problems that arise on poorly configured network devices with other approaches for handling large DNS responses.
\end{abstract}

\section{Introduction}

The Domain Name System (DNS) is a mission critical service for the Internet.
DNS is responsible for translating human-readable domain names into machine-understandable IP
addresses and is used by billions of devices daily. Ensuring that these translations are
correct and not forged is critical to prevent users from being directed to malicious servers
instead of their intended destination. The Domain Name System Security Extensions (DNSSEC) \cite{rfc4033}
provide data integrity by using digital signatures. DNSSEC ensures that the received DNS message
is indeed from a server authorized to respond to the query, and that the message has not been
modified in transit.

Today's DNSSEC uses digital signatures that rely on traditional security assumptions such as factoring and discrete logarithms, which would not resist attacks by a cryptographically relevant quantum computer.  
To continue to provide its intended security guarantees in the face of such threats, DNSSEC must be updated to accommodate quantum-resistant algorithms.
The post-quantum cryptography standardization project of the United States National Institute of Standards and Technology (NIST) announced in July 2022 \cite{NISTround3} three post-quantum digital signatures algorithms to be standardized: CRYSTALS-Dilithium \cite{NISTPQC-R3:CRYSTALS-DILITHIUM20}, Falcon \cite{NISTPQC-R3:FALCON20}, and SPHINCS+ \cite{NISTPQC-R3:SPHINCS+20}.  
All of these selected algorithms have one thing in common: the amount of data transmission required
in order to perform a verification is substantially larger than their non-post-quantum counterparts: both public keys and signatures.
This increase in size can cause substantial issues for pre-existing network protocols; DNS and DNSSEC
are particularly sensitive to this issue.

\paragraph{Constraints on DNS and DNSSEC.}
There is an extremely large quantity of DNS traffic, so DNSSEC must be sufficiently efficient to support this high volume, which leads to the need for highly performant signature verification and, to a somewhat lesser extent, signature generation (signatures are often done offline and then transferred to the servers).
DNS relies primarily on UDP for communicating between servers. UDP has the benefit of being very lightweight
and data efficient, however it has limitations that impact DNS: namely any UDP packet that exceeds 1500 bytes
must be fragmented. UDP fragmentation is fragile and is generally not considered a reliable method for delivering large
messages. With this in mind, accounting for the size of IPv6 headers, it is recommended that the DNS message sizes
should not exceed 1232 bytes \cite{DNSFlagDay20,pqc}. 
As we will note below, for all three of the post-quantum signature algorithms selected by NIST, 1232 bytes is not enough to send both a public key and a signature, as is needed in some parts of DNS.

Admittedly, this 1232 byte limit does not mean that large DNS message cannot in principle be sent. When a DNS response
exceeds 1232 bytes, a truncated response is sent instead indicating to the requester that they should then switch to using TCP instead of UDP.
Unfortunately, a non-trivial number of name servers are estimated to not support TCP communication, preventing them
from sending and receiving large DNS messages \cite{pqc}.

There have been two proposed mechanisms to solve the large DNS message issue \cite{draft-muks-dns-message-fragments-00,draft-song-atr-large-resp-03}, both of which ultimately failed at getting
standardized for use. Both mechanisms moved message fragmentation from the transport layer into the application layer,
thus removing concerns of UDP fragmentation fragility and the lack of support of TCP. If a large DNS message needed to be
sent, both of these mechanisms would split the DNS message into chunks and send each chunk one after the other. 
Fundamentally, both these mechanisms sent many, potentially large, packets, in response to a single request.
There were significant concerns about the impacts these mechanisms would have. 
First, sending a many, potentially large, packets in response to a single request increases the risk and impact of denial of server amplification attacks. 
Second, sending many UDP packets in response to a single UDP request is an unusual behaviour, and some networks are configured to only accept a single UDP response packet to a single UDP request; the rest would trigger ICMP `destination unreachable' packets, leading to concerns about ICMP flooding (which could reduce the utility of ICMP packets in debugging network issues).

Application level fragmentation is not the only solution presented for delivering large messages. Beernink presented in his thesis
the idea of delivering large DNSKEYs out-of-band from DNS. The idea is that when a large DNSKEY is required, such as when using the now defunct round 3 candidate Raindbow \cite{NISTPQC-R3:Rainbow20}, for verification
the requesting server would initiate a HTTP or FTP request to fetch the large key.

\paragraph{Implications for post-quantum DNSSEC.}
When considering which post-quantum algorithms to standardize for DNSSEC, we must consider both the algorithms' operation
performance as well as the sizes of its signatures and public keys. 
Müller et al. \cite{pqc} began this discussion by evaluating the NIST Round 3 
candidates in the context of DNSSEC. They established several requirements for a
scheme to fulfil if it were to be used for DNSSEC signatures. As
noted above, fragmentation is a major concern for DNSSEC
and the recommended maximum DNS response size, including any signatures and
public keys, should not exceed 1232 bytes. However, due to public keys
not needing to be transmitted as often as signatures, larger public keys
may be acceptable. Müller et al. also noted the requirement that
a resolver should be able to validate at least 1000 signatures per second.
The final requirement noted by Müller et al. is that zones should be able to sign 100
records per second.

Müller et al. identified three of the NIST Round 3 candidate algorithms 
that had the potential to fulfill these requirements:
Falcon-512 \cite{NISTPQC-R3:FALCON20}, Rainbow-I$_a$ \cite{NISTPQC-R3:Rainbow20} and RedGeMSS128 \cite{NISTPQC-R3:GeMSS20}. On first
inspection it would appear that Falcon-512 is the clear winner as it is
the only scheme that completely meets the requirements set above,
however, both  Rainbow-I$_a$ and RedGeMSS128 have significantly smaller
signatures sizes which made them appealing: Falcon-512 has a
signature size of 0.7kB whereas the other two schemes have signature
sizes of 66 bytes and 35 bytes respectively. The requirement that both
Rainbow-I$_a$ and RedGeMSS128 failed was that their public keys are 158kB and
375kB respectively, versus Falcon-512's much smaller size of
0.9kB. (Since the 2020 study of Müller et al., both Rainbow and GeMSS have succumbed to cryptanalysis that substantially undermines their claimed security \cite{EC:Beullens21,EPRINT:Beullens22}, and they were not selected by NIST to advance beyond Round 3.)
A conclusion of Müller et al. was that they expect that DNSSEC specification
changes will be required before quantum safe cryptography can be
deployed in order to support larger key sizes.

\subsection{Our contributions}

\begin{table}[t]
\centering
\caption{Resolution times and data transfer sizes for standard DNS (over UDP using TCP fallback) and parallel ARRF in one network scenario.}
\label{tab:intro-stats}
\begin{tabular}{lcc}
\toprule
\textbf{Algorithm}           & \textbf{Standard} & \textbf{Parallel} \\
            & \textbf{DNS} & \textbf{ARRF} \\
\midrule
\multicolumn{3}{c}{\textit{Resolution time (ms) with 10ms latency}} \\
\multicolumn{3}{c}{\textit{and 50 megabytes per second bandwidth}} \\
\midrule
Falcon-512           & 82.11        & 61.96        \\
Dilithium2           & 82.24        & 62.52        \\
SPHINCS+-SHA256-128S & 82.59        & 63.45        \\
RSA 2048 with SHA256 & 41.50        & ---          \\
ECDSA P256           & 47.78        & ---          \\
\midrule
\multicolumn{3}{c}{\textit{Data transfer (bytes)}}         \\
\midrule
Falcon-512           & 3,112        & 2,557        \\
Dilithium2           & 8,623        & 8,367        \\
SPHINCS+-SHA256-128S & 26,073       & 26,140       \\
RSA 2048 with SHA256 & 1,081        & ---        \\
ECDSA P256           & 504          & ---          \\
\bottomrule
\end{tabular}
\end{table}

Given the inherent conflict between the larger public key and signature sizes of post-quantum algorithms and the practical 1232-byte limit on DNS packet size, we revisit fragmentation in hopes of finding a practical way forward.
In this work we propose \protocolnamelong, or \protocolname for short. \protocolname is a \emph{request-based}
lightweight DNS fragmentation solution which removes the fragility of large DNS messages over UDP while
being designed with backwards compatibility in mind. Similarly to previously proposed mechanisms, fragmentation is moved
from the transport layer to the application layer, thus avoiding the fragility of UDP fragmentation.
Whereas previously proposed mechanisms sent several response fragments for a single request, \protocolname
requires that fragments of specific resource records be explicitly requested.
In particular, for large responses, the first response packet is truncated but includes sufficient information to allow the requester to make separate requests for each additional fragment, either in sequential or in parallel (the latter of which we called ``batched \protocolname'').
Our fragmentation approach based on explicit requests for fragments improves both backwards
compatibility and addresses the concern over ICMP flooding. \protocolname is also designed in such a way that
it can be implemented with low impact on existing servers; in fact we were able to implement it as a transparent daemon sitting in front of an \protocolname-unaware requester and resolver at both ends of a DNS lookup request, reducing the
burden of deployment.

To evaluate our approach, we implemented the three post-quantum digital signature algorithms selected by NIST -- specifically, parameter sets Falcon-512, Dilithium2, and SPHINCS+-SHA256-128S -- in BIND using liboqs \cite{SAC:SteMos16}, as well as a daemon implementing \protocolname sitting in front of the requester and resolver, transparently carrying out the \protocolname fragmentation/reassembly.
We were then able to carry out a variety of experiments on a simulated network with different latencies and bandwidth and different fragmentation sizes to evaluate the performance of \protocolname compared to DNS over UDP with TCP fallback, measuring the total resolution time and the amount of data transmitted.

Detailed results across all the various scenarios can be found in \Cref{sec:evaluation}.
\Cref{tab:intro-stats} shows the results for a low-latency (10ms) network scenario, when restricting DNS messages to be at most 1232 bytes.  In this scenario, \protocolname in batched mode (meaning with additional fragments requested in parallel) yields resolution times of approximately 62--63ms for our three post-quantum algorithms, compared to approximately 82ms when using standard DNS over UDP with TCP fallback.  \protocolname is also more data efficient for Falcon-512 and Dilithium2, with the small additional overhead on each \protocolname fragment packet being outweighed by the cost of falling back to TCP and retransmitting the first fragment.

In all our tested scenarios, we found that Falcon-512 performs better than Dilithium2 due to Falcon-512's smaller signatures, suggesting that Falcon-512 may be the most suitable option currently available to be standardized for DNSSEC. 
We did however find that even with the improved performance of post-quantum algorithms in \protocolname compared to standard DNS over UDP with TCP fallback, post-quantum algorithms incurred a performance penalty compared to non-post-quantum algorithms currently in use with DNSSEC (RSA and ECDSA) due to the unavoidable cost of transmitting more data.
Overall, we conclude that \protocolname is a promising option for transitioning to post-quantum DNSSEC: it has less performance degradation compared to standard DNS over UDP with TCP fallback.

It remains to evaluate the backwards compatibility of \protocolname in real-world deployments, where there may be misconfigured network devices or poorly written software that incorrectly handles unrecognized fields.  
We did design \protocolname to avoid some known problems by using EDNS(0) pseudo resource records and using request-based fragmentation rather than responder fragmentation.
Assessing the success of this approach in real-world network scenarios is an important next step.

\section{The Domain Name System}
The Domain Name System is a distributed database primarily responsible for
translating human readable domain names to machine understandable IP addresses.
The DNS is broken up into \emph{zones}, each responsible for a specific level of
granularity of the translation process. Each zone is contains various types
of \emph{resource records} which correspond to \emph{labels}. Resource records can be used
to look up IP addresses associated to domain names, name servers of a zone,
as well as many other types of data.

To assist with explaining how DNS translations are performed, we will suppose
there is a client which wants the IP address for \texttt{\lookupname}. The client
will generally send a query to a caching resolver to handle the rest of the translation
on behalf of the client. Assuming the resolver does not have the answer to the \texttt{\lookupname}
query, it will then query the root name servers for the name servers responsible for \texttt{\lookuptld}
domain names. Once the resolver receives a reply from the root name servers, it will then
query the name servers responsible for \texttt{\lookuptld} for the name servers responsible for \texttt{\lookupname}.
Finally, once the resolver learns of the name servers responsible for \texttt{\lookupname}, it will
query those servers for the IP address associated with \texttt{\lookupname}, and finally receive
and forward the response to the client. The responses to each of the intermediate
queries can be cached to reduce the resolution time and reduce load on name servers.

DNSSEC adds digital signatures to DNS to maintain data integrity. Resource record labels are
not required to be unique, so all resource records of a specified type and a specified label
are grouped together as a RRSet. These RRSets are then signed by a specified digital signature
algorithm, and the signature is stored inside of an RRSIG resource record. The public
key is published to the zone inside of a DNSKEY resource record. There are generally
two types of key pairs generated: Zone Signing Keys (ZSK), and Key Signing Keys (KSK). The
ZSKs are responsible for signing and verifying the resources records in the zone, and the KSKs are
responsible for signing the ZSKs and are what allows the chain of trust to be constructed.

As queries are made from the root servers to its children, and its children's children, eventually
reaching the appropriate name server to answer the query, a chain of trust is constructed. Each zone
that is queried must have a digest of the public KSK being used stored in a delegate signer (DS) record in
its parent's zone, otherwise the public ZSK which is transmitted by the name server cannot be trusted.
The one zone which does not publish a DS record is the root zone, due to its lack of parent. The public
KSK of the root zone must be retrieved out-of-band from DNS; most modern operating systems
have the root zone's public KSK pre-installed, removing the need for the user to fetch and configure the key
themselves.

DNS as original specified only allows for DNS messages of at most 512 bytes over UDP, which quickly became
too small to transport DNS messages, especially with DNSSEC being deployed. Extension Mechanisms for DNS (EDNS(0)) \cite{rfc6891} introduced a way for
resolvers to advertise the maximum sized UDP message they can receive, with a theoretical maximum of $2^{16}$
bytes. In reality, however, UDP/IP fragmentation can pose a significant issue for reliable delivery and thus
the maximum recommended DNS message size over UDP is 1232 bytes \cite{DNSFlagDay20}.

\section{Request-based fragmentation}

As DNS is most reliable with limited size, single packets running over UDP, and given that post-quantum digital signature schemes have public key and signature sizes larger than can be accommodated in that limited size, something must change in order to reliably support post-quantum cryptography in DNSSEC.
In a perfect world, we could simply
send the larger DNS messages with little to no concern of
them arriving. However, UDP
fragmentation can cause significant problems for delivering
large DNS message via UDP. The current solution to solving this
problem is falling back to TCP; however, a non-trivial number of
DNS name servers do not support TCP, and fallback to TCP can also incur a performance penalty. We look to solve this problem
by moving DNS message fragmentation from UDP (transport layer) to
DNS itself (application layer), while addressing concerns raised
to previously proposed mechanisms. 
In this section we present our solution, \protocolnamelong, or \protocolname for short.

\subsection{Resource Record Fragments}
When a DNS message is too large to fit into the maximum advertised
UDP size, some of the message must be omitted while still containing meaningful
information to the requester. We introduce a new type of pseudo-resource
record: Resource Record Fragments (RRFRAGs). Like OPT \cite{rfc6891}, another pseudo-resource
record, RRFRAGs are not explicitly in DNS zones. Rather they are created
only when they are needed. RRFRAGs are designed similarly to the OPT pseudo-resource
record; they use the standard resource record wire format but repurpose some of the
fields. An RRFRAG contains the following fields:

\begin{itemize}
\item \textbf{NAME}: Must always be root (.) to reduce the amount of overhead
required to send a RRFRAG while respecting the generic resource record format.
\item \textbf{TYPE}: Used to identify that this pseudo-resource record is an RRFRAG.
\item \textbf{RRID}: Used to indicate the particular resource record that is
being fragmented. Since labels do not necessarily have distinct resource records
attached to them, this allows a requester to be explicit in its request while not
requiring the responder to remember which particular resource record it fragmented.
The RRID of a particular resource record can be arbitrarily assigned, but must not change.
\item \textbf{CURIDX}: The current index in the byte array of the original
resource record which is being fragmented.
\item \textbf{FRAGSIZE}: The total number of bytes contained in FRAGDATA plus two bytes to account
for the extra space needed for the RRSIZE field. FRAGSIZE has two different meanings depending on the context. If
the RRFRAG is part of a query, then this indicates how large the responding server
should make this particular fragment. If the RRFRAG is part of a response, this field
indicates how much data was sent in this particular fragment.
\item \textbf{RRSIZE}: The size of the original non-fragmented resource record.
This is used by the requester to determine how much data it still needs to request
from the responder in order to reassemble that particular resource record.
\item \textbf{RAGDATA}: The raw bytes of the fragment of the original resource record. In queries
this is always empty. In responses this will contain FRAGSIZE bytes starting at CURIDX.
It is possible for FRAGDATA to contain zero bytes in responses, which we will elaborate
on later.
\end{itemize}

\Cref{fig:rrfrag-mapping} depicts how an RRFRAG maps onto the generic resource record
format. Similar to a DNSKEY resource record where the extra fields
reuqired are inside RDATA, an RRFRAG stores the RRSIZE alongside FRAGDATA inside RDATA.
This was done to handle the case where an implementation which does not support \protocolname
blindly copies RDLENGTH, or in our case FRAGSIZE, bytes into a buffer prior to
branching based on resource record type. 

\begin{table*}
\centering
\arrayrulecolor{black}
\begin{tabular}{r|c|l|c|l}
\multicolumn{1}{l}{}                      & \multicolumn{1}{c}{\begin{tabular}[c]{@{}c@{}}Generic\\Resource Record\\Format\end{tabular}} & \multicolumn{1}{l}{} & \multicolumn{1}{c}{\begin{tabular}[c]{@{}c@{}}RRFRAG\\Format\end{tabular}} &                      \\ 
\hhline{~-~-~}
Variable Bytes                            & {\cellcolor[rgb]{0.722,0.722,0.722}}NAME                                                     & \tikz{\draw[stealth-,thick] (0,0) -- (2,0); }                & {\cellcolor[rgb]{0.733,0.733,0.733}}NAME = `.'                             & 1 Byte               \\ 
\hhline{~-~-~}
2 Bytes                                   & {\cellcolor[rgb]{0.847,0.157,0.541}}TYPE                                                     & \tikz{\draw[stealth-,thick] (0,0) -- (2,0); }                & {\cellcolor[rgb]{0.847,0.157,0.541}}TYPE = RRFRAG                          & 2 Bytes              \\ 
\hhline{~-~-~}
2 Bytes                                   & {\cellcolor[rgb]{0.961,0.596,0.173}}CLASS                                                    & \tikz{\draw[stealth-,thick] (0,0) -- (2,0); }                & {\cellcolor[rgb]{0.961,0.596,0.173}}RRID                                   & 2 Bytes              \\ 
\hhline{~-~-~}
4 Bytes                                   & {\cellcolor[rgb]{0.992,0.776,0.408}}TTL                                                      & \tikz{\draw[stealth-,thick] (0,0) -- (2,0); }                & {\cellcolor[rgb]{0.992,0.776,0.604}}CURIDX                                 & 4 Bytes              \\ 
\hhline{~-~-~}
2 Bytes                                   & {\cellcolor[rgb]{0.973,1,0.259}}RDLENGTH                                                       & \tikz{\draw[stealth-,thick] (0,0) -- (2,0); }                & {\cellcolor[rgb]{0.973,1,0.173}}FRAGSIZE                                   & 2 Bytes              \\ 
\hhline{~-~-~}
\multicolumn{1}{l|}{Up to~$2^{16}$ Bytes} & {\cellcolor[rgb]{0.255,0.992,0.298}}RDATA                                                    & \tikz{\draw[stealth-,thick] (0,0) -- (2,0); }                & {\cellcolor[rgb]{0.255,0.992,0.298}}RRSIZE                                 & 2 Bytes              \\ 
\hhline{~>{\arrayrulecolor[rgb]{0.384,0.765,0.784}}-~>{\arrayrulecolor{black}}-~}
\multicolumn{1}{l|}{}                     & {\cellcolor[rgb]{0.384,0.765,0.784}}                                                         & \tikz{\draw[stealth-,thick] (0,0) -- (2,0); }                & {\cellcolor[rgb]{0.384,0.765,0.784}}FRAGDATA                               & Up to $2^{16}-2$ Bytes  \\
\multicolumn{1}{l|}{}                     & \multirow{-2}{*}{{\cellcolor[rgb]{0.384,0.765,0.784}}}                                       &                      & \multicolumn{1}{l|}{{\cellcolor[rgb]{0.384,0.765,0.784}}}                  &                      \\
\hhline{~-~-~}
\end{tabular}

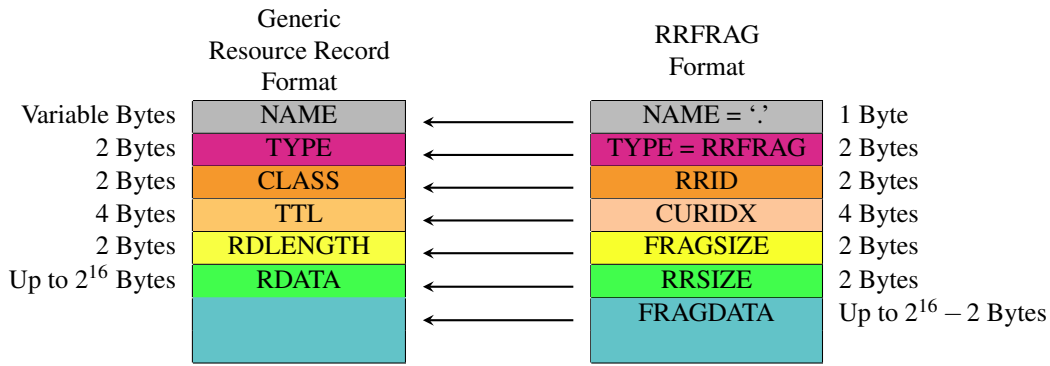
\captionof{figure}{The mapping of the RRFRAG format onto the generic resource record format.}
\label{fig:rrfrag-mapping}
\end{table*}
\subsection{Using RRFRAGs}

When a DNS response is too large to fit in the maximum advertised UDP size, RRFRAGs are
used to split the data across multiple queries with each response's size below the advertised threshold. Resource records
are replaced with RRFRAGs in place. That is to say, that if a resource record being
fragmented is in a particular section of the DNS message, the RRFRAG replacing the resource record will be
inserted into the same section. This is essential so that the original message format, once
all resource records are assembled, will remain intact. It is important to note that
the OPT pseudo-resource record must not be fragmented as it contains important meta
data about the response, such as the DNS cookie. DNS messages that contain RRFRAGs
should send as much data as they are able without surpassing the advertised threshold.

The initial response containing at least one RRFRAG can be considered a ``map'' of the
non-fragmented message. This map is used by the requester to determine what the non-fragmented
DNS message will look like upon reassembly. The requester can now determine what fragments it
is missing in order to complete the original large DNS message, and can now send a new query
for the missing RRFRAGs. It is the responsibility of the requester to specify which resource
records it desires, how large the fragments should be, and where the fragments start. This
is done by adding a RRFRAG for each distinct RRID the requester is requesting a fragment for in the
query's additional section. If the response contains any non-RRFRAG resource records, it
should store them until it is possible to reassemble the entire DNS message.

When the responder sees a query containing a RRFRAG, it just has to construct a standard
DNS response by inserting the corresponding RRFRAGs into the answers section. The Fragdata being
sent is a simple copy of the bytes of the desired resource record starting at CURIDX and ending at
CURIDX + FRAGSIZE. This request/response cycle continues until the requester is able to reassemble the
original large non-fragmented message. 
Note that, after receiving the initial response containing the map, nothing prevents the requester from making the subsequent RRFRAG requests in parallel.

For backwards compatibility reasons, whenever a response
is sent which contains an RRFRAG, the truncated flag (TC) must be set in the DNS message header.

If a requester asks for a fragment which cannot be constructed, such as an RRID which does not map to a
specific resource record, the responder should respond with a return code of FORMERR to indicate that the query
was malformed.

\subsection{Example execution of \protocolname}
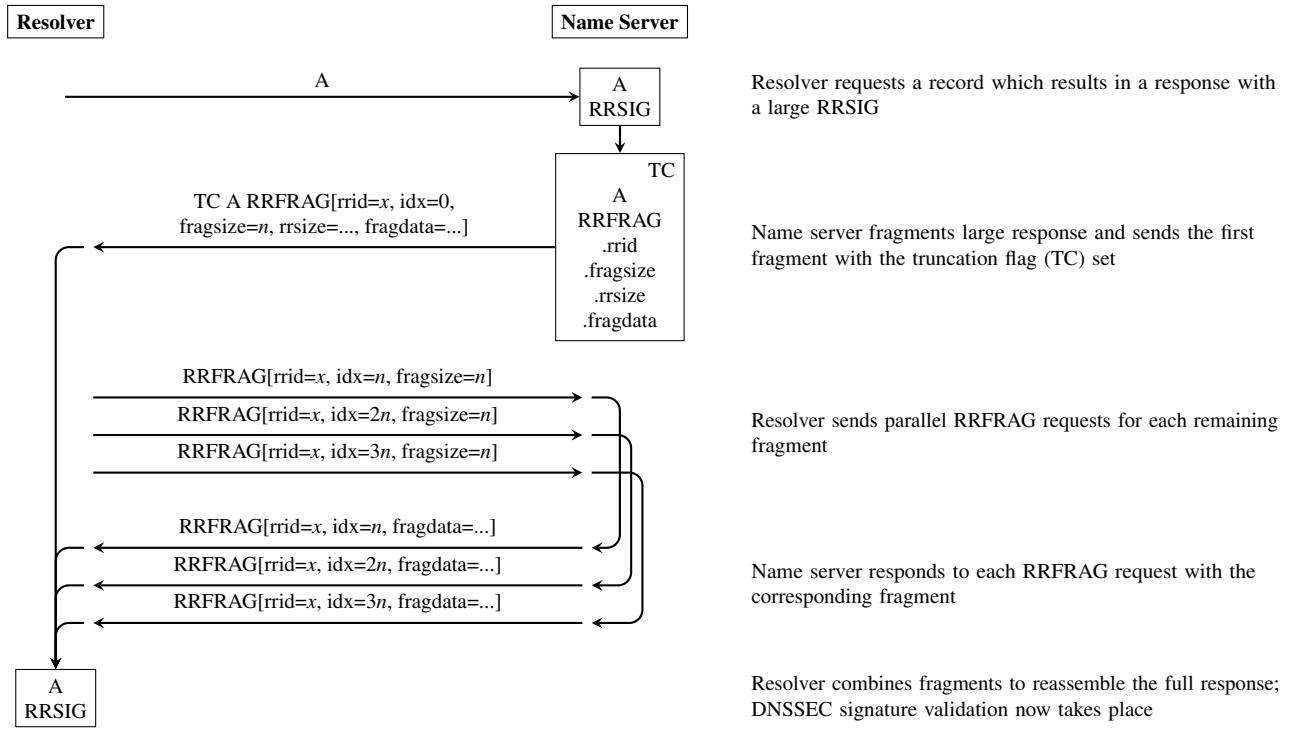
\begin{figure*}[t]
\centering

\begin{tikzpicture}[>=stealth,xscale=1.5,yscale=-1,
	every node/.style={
		font=\footnotesize,
		align=center
	}
]
\node[draw] (LabelResolver) at (0,0) {\textbf{Resolver}};
\node[draw] (LabelNameServer) at (5,0) {\textbf{Name Server}};
\node (req1) at (0,1) {};
\node[draw] (resp1) at (5, 1) {A\\RRSIG};
\draw[->,thick] (req1) -- node[above] {A} (resp1);

\node[align=left,text width=7cm] at (8.5, 1) {Resolver requests a record which results in a response with a large RRSIG};

\node[draw] (resp2) at (5, 3) {\phantom{RRFRAG}TC\\A\\RRFRAG\\.rrid\\.fragsize\\.rrsize\\.fragdata};
\draw[->,thick] (resp1)--(resp2);

\node[align=left,text width=7cm] at (8.5, 3) {Name server fragments large response and  sends the first fragment with the truncation flag (TC) set};

\node (req2) at (0.25,3) {};
\draw[->,thick] (resp2) -- node[above]{TC A RRFRAG[rrid=$x$, idx=0, \\ fragsize=$n$, rrsize=..., fragdata=...]} (req2);

\node (req3) at (0.25, 5) {};
\node (req4) at (0.25, 5.5) {};
\node (req5) at (0.25, 6) {};
\node (resp3) at (4.75, 5) {};
\node (resp4) at (4.75, 5.5) {};
\node (resp5) at (4.75, 6) {};

\draw[->,thick] (req3) -- node[above]{RRFRAG[rrid=$x$, idx=$n$, fragsize=$n$]} (resp3);
\draw[->,thick] (req4) -- node[above]{RRFRAG[rrid=$x$, idx=$2n$, fragsize=$n$]} (resp4);
\draw[->,thick] (req5) -- node[above]{RRFRAG[rrid=$x$, idx=$3n$, fragsize=$n$]} (resp5);

\node[align=left,text width=7cm] at (8.5, 5.5) {Resolver sends parallel RRFRAG requests for each remaining fragment};

\node (resp6) at (4.75,7) {};
\node (resp7) at (4.75,7.5) {};
\node (resp8) at (4.75,8) {};
\node (req6) at (0.25,7) {};
\node (req7) at (0.25,7.5) {};
\node (req8) at (0.25,8) {};

\draw[->,thick,rounded corners=5pt] (4.75,5) -- (5,5) -- (5,7) -- (4.75,7);
\draw[->,thick,rounded corners=5pt] (4.75,5.5) -- (5.1,5.5) -- (5.1,7.5) -- (4.75,7.5);
\draw[->,thick,rounded corners=5pt] (4.75,6) -- (5.2,6) -- (5.2,8) -- (4.75,8);
\draw[->,thick] (resp6) -- node[above]{RRFRAG[rrid=$x$, idx=$n$, fragdata=...]} (req6);
\draw[->,thick] (resp7) -- node[above]{RRFRAG[rrid=$x$, idx=$2n$, fragdata=...]} (req7);
\draw[->,thick] (resp8) -- node[above]{RRFRAG[rrid=$x$, idx=$3n$, fragdata=...]} (req8);

\node[align=left,text width=7cm] at (8.5, 7.5) {Name server responds to each RRFRAG request with the corresponding fragment};

\node[draw] (assemble) at (0, 9) {A \\ RRSIG};

\draw[->,thick,rounded corners=5pt] (0.25,3) -- (0,3) -- (assemble.north);
\draw[->,thick,rounded corners=5pt] (0.25,7) -- (0,7) -- (assemble.north);
\draw[->,thick,rounded corners=5pt] (0.25,7.5) -- (0,7.5) -- (assemble.north);
\draw[->,thick,rounded corners=5pt] (0.25,8) -- (0,8) -- (assemble.north);

\node[align=left,text width=7cm] at (8.5, 9) {Resolver combines fragments to reassemble the full response; DNSSEC signature validation now takes place};
\end{tikzpicture}

\caption{An example execution of \protocolname}
\label{fig:RRFRAGexample}
\end{figure*}

To better solidify how \protocolname works, we will now work through an example DNS query whose
response is larger than the MTU. This example has had some details abstracted away and should
not be used in place of the above specification when implementing \protocolname. 
\Cref{fig:RRFRAGexample} illustrates our example execution. This example begins at the last
stage of name resolution for the query ``\lookupname''. We have two parties: the resolver
making the DNSSEC-enabled query for \lookupname., and the \lookupname.
name server.

First the resolver makes a standard request for the A record
and its associated RRSIG. Upon receiving the request, the
resolver observes that the DNS response is too large to fit within the confines of the MTU, and thus replaces the large RRSIG with an RRFRAG.
This RRFRAG will contain as much of the original RRSIG as possible, and will inform the resolver
how much of the original RRSIG is missing. Once the resolver receives the DNS response, it will
copy both the entire A record as well as the RRFRAG and allocating
enough space for the rest of the missing record. The resolver will then send another DNS query,
but this time asking for an RRFRAG and sending its own RRFRAG indicating the next range of data it needs.
Once the name server receives the RRFRAG query, it will use the RRFRAG in the additional
section to determine the starting position and size of the fragment of the original RRSIG is being
requested. The name server will construct a new DNS response containing the rest of our missing RRSIG
inside of an RRFRAG and send the new response to the resolver. Finally
the resolver will copy the newly received RRFRAG into its state, reassemble the original RRSIG, and
finally reconstruct the original large DNS response. DNSSEC validation
now takes place, and if verification is successful the records are cached by the resolver.

\subsection{Caching and DNSSEC Considerations}
RRFRAGs themselves should never be cached. Once a DNS message is reassembled, and its DNSSEC authenticaiton validated if appropriate,
then then non-fragmented resource records may be cached.
If RRFRAGs could be cached, this would allow for malicious data to be accepted prior
to validation. Caching complete resource records as opposed to RRFRAGs also allows
for intermediate resolvers to send different fragment sizes than they originally received
which allows for more flexibility to handle varying advertised UDP sizes. 

\section{Evaluation}\label{sec:evaluation}
In this section we evaluate the performance of post-quantum signature algorithms in DNSSEC without and with our request-based fragmentation technique \protocolname.

\subsection{Experiment setup}

\paragraph{Algorithms.}
The algorithms we selected for the experiment are level-1 (128-bit-security) parameter sets of the three algorithms selected for standardization by NIST
at the end of round 3: Falcon-512, Dilithium2-AES, and SPHINCS+-SHA256-128S.
We also include results for RSA 2048 with SHA256 and ECDSA P256 for the sake of comparison.

\paragraph{Adding post-quantum algorithms to BIND.}
We evaluate these algorithms both using DNSSEC as defined today, as well as with \protocolname.
To perform this evaluation we used Internet Systems Consortium's BIND9 9.17.9 \cite{bind} as our
DNS server software. We then added support for the three selected algorithms to BIND9 using
Open Quantum Safe's liboqs 0.7.1 and OpenSSL 1.1.1l fork \cite{SAC:SteMos16,OQS}.
To construct a test network environment,
we used Docker and Docker's built in networking as well as Linux's `\texttt{tc}' (traffic control) to simulate network bandwidth
and latency. 

\paragraph{Daemon implementing \protocolname.}
Rather than implementing \protocolname directly into BIND9, we constructed a daemon which intercepts all
incoming and outgoing network traffic and implements \protocolname transparently for both
the resolver and all name servers. We used libnetfilter-queue 1.0.3-1 to intercept packets.

We will now describe how the daemon behaves. When the machine acting as the name server receives a DNS query, the daemon on the name server's side will modify the maximum advertised
UDP message size to the maximum value of 65355 bytes.\footnote{Modifications to BIND9 were required as the maximum DNS message size BIND9 supports is 4096}
The daemon then sends the message to the DNS software,
which responds with a UDP message up to 65355 bytes. 
The daemon on the name server side receives
this response and copies the entire message into its state.
It outputs a response that is either the original message, if it fits within the requester's maximum UDP message size, or the first fragment if fragmentation is required.
Whenever
a fragment is requested in the future the daemon will use its state if possible rather than sending the request to
the DNS software. 

On the side of the DNS resolver, there is another copy of the daemon which intercepts incoming DNS responses and processes them before passing them on to the DNS resolver.
When the resolver-side daemon receives a DNS response containing an RRFRAG,
the daemon will intercept the message. The daemon will create a state for that individual transaction containing
the metadata provided by the initial response's map and copy any data included into the state. The daemon will
then execute \protocolname and request the remaining fragments until the entire message can be reconstructed, at which point in time the daemon transparently sends the reconstructed message to the DNS resolver software.

\paragraph{DNS network design.}
We construct a simple DNS network consisting of a client, a resolver, and a name server each running in their own Docker container on the same machine.
The name server zone contains 1000 `A' records, each with a unique label and signature. We query for each of these A records and measure the total resolution time for each one. The zone also contains 1 `primer' name record. We first query for this primer resource record so that our resolver has the DNSKEYs and NS records of our test domain, which means that we can evaluate \protocolname's effect on an individual query.
To model the worst case response size, we disabled `minimal responses', and as such each response will contain 1 question, 1 A record, 1 NS record, 1 SOA record, and 3 RRSIGs.
We use `\texttt{dig}' to issue each query and measure the total resolution time of said query.

We evaluated using the following four network conditions:
\begin{itemize}
\item low bandwidth, low delay: 10ms of delay and 128 kilobytes per second bandwidth;
\item high bandwidth, low delay: 10ms of delay and 50 megabytes per second bandwidth;
\item moderate bandwidth, high delay: 100ms of delay and 50 megabytes per second bandwidth; and
\item ideal network: no delay, unlimited bandwidth (the only cost being processing the messages).
\end{itemize}

All experiments were run on a c5.2xlarge Amazon Web Services instance which provides 8 cores of a 3GHz Intel Xeon Platinum 8124M and 16 gigabytes of RAM.

\subsection{Algorithm performance}
To put the network results in context, it is important to understand the performance of the verification function of
each of the algorithms. We use the Open Quantum Safe OpenSSL fork's \texttt{speed} command to measure each algorithm's signing
and verification performance and report the results in \Cref{tab:openssl-speed}.

\begin{table}[t]
\caption{Algorithm runtime measured using OQS-OpenSSL Speed}
\label{tab:openssl-speed}
\centering
\begin{tabular}{lrr}
\toprule
\textbf{Algorithm}            & \textbf{Sign (ms)} & \textbf{Verify (ms)} \\
\midrule
Falcon-512            & 0.2810    & 0.0438      \\
Dilithium2           & 0.0753    & 0.0268      \\
SPHINCS+-SHA256-128S & 373.1\hphantom{000}     & 1.36\hphantom{00}       \\
RSA 2048 with SHA256 & 0.6019    & 0.1772      \\
ECDSA P256           & 0.0219    & 0.0677    \\
\bottomrule
\end{tabular}\end{table}

\subsection{Post-quantum with standard DNSSEC}
In this section we measured how the post-quantum algorithms perform if they are deployed in DNSSEC as
it is currently specified, under two scenarios and five different network conditions. We first
measured how the algorithms would perform with a maximum UDP size of 1232. For messages larger than
1232 bytes, the DNS servers will fall back to TCP. The second scenario is the exclusive use of UDP for DNS communication, which
provides an idealized view of the best case performance we can achieve using a particular algorithm; in this scenario, responses larger than the maximum advertised UDP message size will be fragmented \emph{by the responder}, resulting in multiple UDP packets being sent in response to a single UDP packet request.
\Cref{tab:no-arrf-resolution} shows the average resolution
times with standard deviation for the various network conditions. RSA 2048 with SHA256 and ECDSA P256 only have results recorded for standard DNS as the signatures
of these algorithms are small enough to ensure they can fit in a single DNS message without fragmentation.

\begin{table}[t]
\centering
\caption{Mean resolution times (and standard deviation) in milliseconds for DNS without ARRF}
\label{tab:no-arrf-resolution}
\begin{tabular}{lcc}
\toprule	
\textbf{Algorithm} & \textbf{Standard DNS} & \textbf{DNS using only UDP} \\
\midrule
\multicolumn{3}{c}{\textit{10ms of latency and 128 kilobytes per second bandwidth}} \\ \midrule
Falcon-512                   & 107.3 $\pm$ 1.786        & 61.52 $\pm$ 2241          \\
Dilithium2                   & 147.9 $\pm$ 1.478        & 102.0 $\pm$ 1.898         \\
SPHINCS+         & 275.4 $\pm$ 2.114        & 229.4 $\pm$ 2.040         \\
RSA 2048         & 52.20 $\pm$ 1.242        & ---                       \\
ECDSA P256                   & 47.78 $\pm$ 1.949        & ---                       \\ \midrule
\multicolumn{3}{c}{\textit{10ms of latency and 50 megabytes per second bandwidth}}  \\ \midrule
Falcon-512                   & 82.11 $\pm$ 2.331        & 40.56 $\pm$ 2.115         \\
Dilithium2                   & 82.24 $\pm$ 2.216        & 40.77 $\pm$ 2.251         \\
SPHINCS+         & 82.59 $\pm$ 2.096        & 41.16 $\pm$ 2.192         \\
RSA 2048         & 41.50 $\pm$ 2.157        & ---                       \\
ECDSA P256                   & 47.49 $\pm$ 1.919        & ---                       \\ \midrule
\multicolumn{3}{c}{\textit{100ms of latency and 50 megabytes per second bandwidth}}          \\ \midrule
Falcon-512                   & 802.1 $\pm$ 2.115        & 401.6 $\pm$ 1.991         \\
Dilithium2                   & 802.4 $\pm$ 2.032        & 401.5 $\pm$ 1.962         \\
SPHINCS+         & 802.5 $\pm$ 1.940        & 401.9 $\pm$ 2.021         \\
RSA 2048         & 401.3 $\pm$ 2.022        & ---                       \\
ECDSA P256                   & 401.2 $\pm$ 2.176        & ---                       \\ \midrule
\multicolumn{3}{c}{\textit{0ms of latency and unlimited bandwidth}}                 \\ \midrule
Falcon-512                   & 2.480 $\pm$ 3.884        & 1.1222 $\pm$ 2.034        \\
Dilithium2                   & 2.282 $\pm$ 3.318        & 1.240 $\pm$ 2.156         \\
SPHINCS+         & 2.38 $\pm$ 3.500         & 1.176 $\pm$ 1.935         \\
RSA 2048         & 1.672 $\pm$ 3.046        & ---                       \\
ECDSA P256                   & 1.567 $\pm$ 2.711        & ---                       \\
\bottomrule
\end{tabular}
\end{table}

\subsection{Post-quantum with \protocolname}
In this section we evaluate how each of the algorithms perform when using two different flavours of \protocolname. First, we consider a ``sequential"
version. This version sends a request, receives a response, then looks what it needs to request and sends another request. This process is repeated until
the entire message is received. Next, we consider a ``parallel'' version where once the first response is received the name server sends all of the requests for the remaining
fragments at once, essentially parallelizing the \protocolname mechanism. We consider several
scenarios where the maximum DNS message size varies across all of the various network conditions described above.

Our daemon implementation is a prototype; with that in
mind, it is important to understand the raw overhead that the daemon incurs. By setting the maximum DNS message size to be larger than any response (say, 65355 bytes),
we can see how much of a cost we are paying just by having the proof of concept daemon involved. We then also evaluate what we would expect
most operators would use as their maximum DNS message size of 1232 bytes. In order to see how \protocolname scales, we also provide
some smaller maximum DNS message sizes of 512 (the minimum DNS message size that must be supported) and 256 bytes.
\Cref{tab:seq-resolution} shows the measured mean resolution time in milliseconds for the daemon running in sequential mode for the various network conditions measured, and \Cref{tab:bat-resolution} contains the results for the parallel daemon. Figures \ref{fig:scatter-10ms-128kbps}, \ref{fig:scatter-10ms-50mbps}, \ref{fig:scatter-100ms-50mbps}, and \ref{fig:scatter-0ms-unlimited} illustrate all measured resolution times for standard DNS and DNS using \protocolname for all network conditions.
\begin{table*}[t]
\centering
\caption{Mean resolution times (with standard deviation) with ARRF using daemon in sequential mode}
\label{tab:seq-resolution}
\begin{tabular}{lcccc}
\toprule
\multirow{3}{*}{\textbf{Algorithm}} & \multicolumn{4}{c}{\textbf{ARRF in sequential mode}} \\
& \multicolumn{4}{c}{\textbf{Resolution times (ms) for each maximum message size}} \\
 &
  \textbf{65355 bytes} &
  \textbf{1232 bytes} &
  \textbf{512 bytes} &
  \textbf{256 bytes} \\ \midrule
\multicolumn{5}{c}{\textit{10ms of latency and 128 kilobytes per second bandwidth}} \\ \midrule
Falcon-512 &
  62.61~$\pm$~2.052 &
  84.414~$\pm$~1.451 &
  148.5~$\pm$~1.587 &
  275.8~$\pm$~1.738 \\
Dilithium2 &
  103.2~$\pm$~1.753 &
  231.7~$\pm$~1.841 &
  422.7~$\pm$~2.409 &
  803.9~$\pm$~1.344 \\
SPHINCS+-SHA256-128S &
  230.7~$\pm$~1.879 &
  635.1~$\pm$~2.088 &
  1271~$\pm$~1.963 &
  2480~$\pm$~1.916 \\ \midrule
\multicolumn{5}{c}{\textit{10ms of latency and 50 megabytes per second bandwidth}} \\ \midrule
Falcon-512 &
  41.77~$\pm$~2.135 &
  62.07~$\pm$~2.278 &
  122.5~$\pm$~2.197 &
  243.0~$\pm$~2.269 \\
Dilithium2 &
  41.91~$\pm$~2.108 &
  162.9~$\pm$~2.240 &
  343.8~$\pm$~1.899 &
  705.6~$\pm$~2.379 \\
SPHINCS+-SHA256-128S &
  42.45~$\pm$~2.160 &
  424.7~$\pm$~1.811 &
  1028~$\pm$~2.465 &
  2173~$\pm$~2.123 \\ \midrule
\multicolumn{5}{c}{\textit{100ms of latency and 50 megabytes per second bandwidth}} \\ \midrule
Falcon-512 &
  401.97~$\pm$~2.060 &
  601.1~$\pm$~2.865 &
  1203~$\pm$~1.912 &
  2404~$\pm$~1.123 \\
Dilithm2 &
  402.1~$\pm$~2.005 &
  1604~$\pm$~1.754 &
  3405~$\pm$~2.113 &
  7008~$\pm$~1.708 \\
SPHINCS+-SHA256-128S &
  402.7~$\pm$~1.957 &
  4207~$\pm$~2.166 &
  10210~$\pm$~1.843 &
  21620~$\pm$~1.440 \\ \midrule
\multicolumn{5}{c}{\textit{0ms of latency and unlimited bandwidth}} \\ \midrule
Falcon-512 &
  \multicolumn{1}{l}{1.644~$\pm$~2.334} &
  \multicolumn{1}{l}{1.992~$\pm$~2.594} &
  \multicolumn{1}{l}{2.172~$\pm$~2.361} &
  \multicolumn{1}{l}{2.668~$\pm$~2.606} \\
Dilithium2 &
  \multicolumn{1}{l}{1.804~$\pm$~2.641} &
  \multicolumn{1}{l}{2.344~$\pm$~2.495} &
  \multicolumn{1}{l}{2.932~$\pm$~2.184} &
  \multicolumn{1}{l}{4.176~$\pm$~1.291} \\
SPHINCS+-SHA256-128S &
  \multicolumn{1}{l}{1.992~$\pm$~2.408} &
  \multicolumn{1}{l}{3.564~$\pm$~1.460} &
  \multicolumn{1}{l}{5.692~$\pm$~2.243} &
  \multicolumn{1}{l}{5.673~$\pm$~2.389} \\
  \bottomrule
\end{tabular}
\end{table*}

\begin{table*}[t]
\centering
\caption{Mean resolution times (with standard deviation) with ARRF using daemon in parallel mode}
\label{tab:bat-resolution}
\begin{tabular}{lcccc}
\toprule
\multirow{3}{*}{\textbf{Algorithm}} & \multicolumn{4}{c}{\textbf{ARRF in parallel mode}} \\
 & \multicolumn{4}{c}{\textbf{Resolution times (ms) for each maximum message size}} \\
 &
  \textbf{65355 bytes} &
  \textbf{1232 bytes} &
  \textbf{512 bytes} &
  \textbf{256 bytes} \\ \midrule
\multicolumn{5}{c}{\textit{10ms of latency and 128 kilobytes per second bandwidth}} \\ \midrule
Falcon-512 &
  62.80~$\pm$~2.161 &
  84.68~$\pm$~1.765 &
  86.15~$\pm$~2.296 &
  89.50~$\pm$~2.120 \\
Dilithium2 &
  103.1~$\pm$~1.855 &
  127.9~$\pm$~1.551 &
  132.9~$\pm$~2.038 &
  142.7~$\pm$~2.024 \\
SPHINCS+-SHA256-128S &
  230.7~$\pm$~1.908 &
  262.9~$\pm$~2.050 &
  279.7~$\pm$~1.720 &
  311.6~$\pm$~2.070 \\ \midrule
\multicolumn{5}{c}{\textit{10ms of latency and 50 megabytes per second bandwidth}} \\ \midrule
Falcon-512 &
  41.62~$\pm$~2.060 &
  61.96~$\pm$~2.140 &
  62.14~$\pm$~2.343 &
  62.16~$\pm$~2.156 \\
Dilithium2 &
  41.02~$\pm$~2.170 &
  62.52~$\pm$~2.240 &
  62.96~$\pm$~2.590 &
  62.45~$\pm$~2.590 \\
SPHINCS+-SHA256-128S &
  42.35~$\pm$~2.164 &
  63.45~$\pm$~2.241 &
  64.44~$\pm$~1.865 &
  66.808~$\pm$~2.247 \\ \midrule
\multicolumn{5}{c}{\textit{100ms of latency and 50 megabytes per second bandwidth}} \\ \midrule
Falcon-512 &
  400.6~$\pm$~1.965 &
  601.1~$\pm$~2.212 &
  601.2~$\pm$~2.208 &
  601.7~$\pm$~2.168 \\
Dilithm2 &
  400.9~$\pm$~2.044 &
  601.7~$\pm$~2.271 &
  601.7~$\pm$~2.209 &
  602.4~$\pm$~1.947 \\
SPHINCS+-SHA256-128S &
  401.5~$\pm$~2.145 &
  602.4~$\pm$~1.870 &
  603.4~$\pm$~1.638 &
  605.5~$\pm$~2.3638 \\ \midrule
\multicolumn{5}{c}{\textit{0ms of latency and unlimited bandwidth}} \\ \midrule
Falcon-512 &
  \multicolumn{1}{l}{1.224~$\pm$~2.428} &
  \multicolumn{1}{l}{1.471~$\pm$~2.250} &
  \multicolumn{1}{l}{1.650~$\pm$~2.310} &
  \multicolumn{1}{l}{1.769~$\pm$~2.520} \\
Dilithium2 &
  \multicolumn{1}{l}{1.185~$\pm$~2.052} &
  \multicolumn{1}{l}{1.698~$\pm$~2.365} &
  \multicolumn{1}{l}{1.875~$\pm$~2.010} &
  \multicolumn{1}{l}{2.496~$\pm$~1.871} \\
SPHINCS+-SHA256-128S &
  \multicolumn{1}{l}{1.436~$\pm$~2.143} &
  \multicolumn{1}{l}{2.406~$\pm$~1.876} &
  \multicolumn{1}{l}{3.461~$\pm$~1.618} &
  \multicolumn{1}{l}{5.673~$\pm$~2.389} \\
  \bottomrule
\end{tabular}
\end{table*}

\subsection{Data transmission}

In order to understand the full implications of deploying \protocolname, we must also consider
the amount of data transmitted compared to that of the DNS as it is currently standardized.
\Cref{tab:data-transmitted} shows the total number of bytes required to transmit a complete DNS message signed with
Falcon-512, Dilithium2, and SPHINCS-SHA256-128S both with and without \protocolname deployed.

\begin{table}[t]
\centering
\caption{Total data transmitted when performing a DNS lookup}
\label{tab:data-transmitted}
\begin{tabular}{lcccc}
\toprule
\multirow{5}{*}{\textbf{Algorithm}} &
  \multicolumn{4}{c}{\textbf{Bytes transmitted during DNS lookup}} \\
 &
    & \multicolumn{3}{c}{\textbf{ARRF}} \\
 & \textbf{Standard} & 
  \multicolumn{3}{c}{\textbf{maximum message size}} \\
                     & \textbf{DNS}  & \textbf{1232}   & \textbf{512}   & \textbf{256}   \\ 
                     &   & \textbf{bytes}   & \textbf{bytes}   & \textbf{bytes}   \\ \midrule
Falcon-512           & 3,112                & 2,557  & 2,947  & 3,637  \\
Dilithium2           & 8,623                & 8,367  & 9,402  & 11,322 \\
SPHINCS+ & 26,073               & 26,140 & 29,620 & 36,175 \\
\bottomrule
\end{tabular}
\end{table}

\subsection{Results}

\paragraph{Resolution times for standard DNS without \protocolname.}
When considering standard DNS, RSA and ECDSA have the shortest resolution times with the best performing post-quantum algorithm being twice as
slow across all network conditions. This is due to the response sizes being too large for a single UDP packet, causing it to
be truncated and thus effectively making the initial query a wasted trip. The resolver must then fall back to the less performant 
TCP protocol to complete the lookup. When standard DNS using only UDP (with name-server-based fragmentation) is used, ECDSA and RSA only beat Falcon-512 and Dilithium2 when bandwidth was
restricted to 128 kilobytes per second; this is likely due to the verification functions of Falcon-512 and Dilithium2 being more efficient than ECDSA and RSA.

\paragraph{Basic overhead of \protocolname daemon.}
When considering the cases where the \protocolname daemon is running, but not actively fragmenting resource records, we see
comparable performance to standard DNS using only UDP. When comparing the post-quantum algorithms on standard DNS using only UDP versus
the \protocolname daemon using a maximum message size of 65355 bytes, we see a minimal overhead never exceeding 1.25 ms. Given that this is the overhead for our prototype daemon running as a separate process, we conclude that \protocolname itself has very low overhead when
fragmentation is not required.

\paragraph{Parallel versus sequential \protocolname.}
When the \protocolname daemon is fragmenting resource records, we see that the parallel daemon has a performance improvement of approximately 20\% over TCP for all algorithms and all maximum messages sizes. This is due
to the parallel nature of the parallel daemon effectively only paying the latency cost once after receiving the initial response, whereas TCP has a limited sized window restricting its parallelization, which cases the latency cost to be paid more times compared to the unlimited parallelization of parallel \protocolname.
The sequential daemon even outperforms TCP for
Falcon-512 with a maximum messages size of 1232 bytes across all tested network conditions. This is due to the Falcon-512 signed response only requiring
one additional round trip to reassemble the message, whereas the TCP fallback needs to receive the entire message from scratch (it cannot make use of the truncated response returned in the UDP response).

The sequential daemon performs worse in all other cases and is greatly affected by increased latency. This is due to the sequential daemon needing to wait for each request to be fulfilled before requesting the next piece, and TCP being able to achieve some parallelism due to its sliding window.

In the scenarios with latency and bandwidth restrictions, we see that, as the maximum message size is reduced, parallel \protocolname scales very nicely due to parallelizing the requests, whereas sequential \protocolname scales roughly by the factor that the maximum message size is reduced by.

\paragraph{Post-quantum versus non-post-quantum.}
When comparing post-quantum to non-post-quantum algorithms, Falcon-512 comes the closest to RSA and ECDSA in all constrained network scenarios, but is still slower
despite the efficient verification function. Falcon-512 is affected primarily by bandwidth and is 60\% slower than RSA and 76\% slower than ECDSA in the 128 kilobytes per second scenario even when using parallel \protocolname. If bandwidth is not a concern, then Falcon-512 performs better, but is still 49\% slower than both RSA and  ECDSA in both scenarios with 50 megabytes per second bandwidth. Unsurprisingly, Dilithium2 and SPHINCS+-SHA256-128S perform far worse than Falcon-512 and the non-post-quantum algorithms; roughly 1.5 and 3 times slower than Falcon-512 when using parallel \protocolname, and even worse when using sequential \protocolname.

\paragraph{Data overhead.}
When DNS messages sizes are at the recommended size of 1232 bytes, we can see that \protocolname actually uses less data to transmit
a DNSSEC response signed with Falcon-512 and Dilithium2. This is due to how DNS handles switching to TCP, essentially causing the three-way TCP handshake to turn into a five-way handshake, which we now explain.
First the resolver sends a UDP request to the name server. The name server then sends a response identical to the request and marks the response as truncated. 
The resolver switches over to TCP and performs the standard
TCP three-way handshake. TCP also sends an acknowledgement packet for each packet the requester receives, essentially offsetting the fragment requests in \protocolname.
With these factors, combined with UDP packet headers being 12 bytes smaller than those of TCP, \protocolname allows efficient communication for both Falcon-512 and Dilithium2. 

However, TCP becomes more data efficient compared to \protocolname once many fragments are requested and sent, such as for SPHINCS+-SHA256-128S. Due to maintaining backwards
compatibility, \protocolname must surround all requests and responses inside of a DNS message and all fragments inside of an RRFRAG. TCP, on the other hand, is a stream which only
sends a single DNS message header and sends the raw resource records themselves rather than sending the extra bytes that RRFRRAGs require. As mentioned earlier TCP sends acknowledgement
packets for each TCP packet received. These acknowledgements are smaller than a UDP packet containing an \protocolname request. The size difference depends on how many RRFRAGs are
being requested, but the most common \protocolname request in our experiments was 60 bytes including UDP, IP, and DNS message headers, and the largest request being 75 bytes, whereas
TCP's acknowledgement packets are 52 bytes in size.  If a DNS message is quite large, as is the case with SPHINCS+-SHA256-128S signed messages, these small savings end up
making up for wasting the initial UDP request.

\section{Discussion}

Having seen the results of the experiments, we now discuss \protocolname and consider whether if it is a viable solution for sending large DNS message.

\subsection{Performance}
Parallel \protocolname is by far the most performant solution for larger responses, beating out TCP fallback in all cases despite how many requests
and responses are required to transmit the original large DNS message. Sequential \protocolname also outperforms TCP in cases where messages are only slightly larger than what can fit in a single UDP packet. 
However, parallel \protocolname's performance does not come for free. On
a busy resolver these parallel requests could eat up available bandwidth quite quickly and could potentially
overwhelm middle boxes. We hypothesize that a production-ready version of \protocolname would have a maximum
window size similar to TCP in an effort to reduce request flooding, and therefore performance would lie somewhere between the ideal version of parallel \protocolname and TCP. 
Despite there not being considerable differences between DNS with only UDP
and the \protocolname daemon running but not fragmenting, there are likely optimizations, such as multithreading, that can be made to the daemon.
If \protocolname was integrated directly into DNS software, it would also increase efficiency.
We leave experimenting and evaluating these potential optimizations as well as evaluating window sizes as future work.

\subsection{Backwards Compatibility}
As DNS is a distributed system managed by many different entities, in any deployment there will be requesters and name servers which do not understand \protocolname. 
We now consider what happens in two such scenarios: when the requester implements \protocolname but the responder does not, and when the requester
does not implement \protocolname but the responder does. We also discuss the impact \protocolname has on middle boxes.

\paragraph{Requester implements \protocolname but responder does not.}
When a requester which supports \protocolname receives a response from a name server
which does not support \protocolname, it will, as per the current DNS specifications, receive a truncated DNS message with
the TC flag set. It can then gracefully fallback to TCP and retry the query, therefore
maintaining backwards compatibility.

\paragraph{Requester does not implement \protocolname but responder does.}
Since the requester does not actually indicate its support of \protocolname, it may appear 
at first glance that \protocolname may cause issues when the requester receives a response
containing an RRFRAG, as it will not be able to understand what an RRFRAG is, nor what
it should do with it. Fortunately, older resolvers ignore unknown resource record types,
so they will gracefully fallback to repeating the request over TCP as they will see that the TC flag is set.
This results
in no additional round trips compared to if \protocolname was not being used.

\paragraph{Middle box support.}
By fragmenting at the DNS level, we should ensure that the majority of middle boxes
will not cause issues for \protocolname. From a middle box's perspective (even one unaware of \protocolname), all messages
sent using \protocolname look like standard DNS messages which should not require any state
to be properly routed. However, if there exist middle boxes which look inside DNS messages
and view the types of the message's resource records, the new RRFRAG type could potentially
cause those middle boxes to reject the message. Additional work would be required to
determine if there are middle boxes with that behaviour, and how widespread they are.

\subsection{Security Considerations}

\paragraph{Denial of service attacks.}
\protocolname is designed to not increase the scope of DoS attacks. Since fragments
must be explicitly requested, a querier can reject any fragments it is not expecting (unlike responder-based fragmenting).
When combined with DNS cookies, off-path attacks become infeasible. An adversary who
is on-path could modify the values in responses which contain RRFRAGs, which could cause
a querier to ask for fragments which do not exist. Middle boxes could also inject
malicious data into individual RRFRAG's FRAGDATA fields. If DNSSEC is used, then this
will cause the validation to eventually fail. This is acceptable as this validation failure,
although denying service, is no worse than DNS without \protocolname deployed (where a middle box adversary simply modifies the body or signature of a DNSSEC response).
\protocolname also limits the impact of amplification DoS attacks as it restricts the
response sizes and each response needs a corresponding request. If a response arrives
with the wrong id or DNS cookie, it should be discarded.

\paragraph{DNS cache poisoning.}
Since RRFRAGs themselves should not be cached, DNS cache poisoning is no more of a
concern than it is in traditional DNS. If DNSSEC is used, then DNS cache poisoning
is not a concern assuming a secure algorithm is used.

\paragraph{Memory exhaustion attacks.}
\protocolname as specified is susceptible to memory exhaustion attacks. Although
DNS cookies make this less of a concern for off-path adversaries, there is nothing
stopping an on-path adversary from changing the RRSIZE fields in the initial response.
Since the requester uses this initial response as a map without any validation thereof, an adversary
could insert many RRFRAGs
advertising they are fragments of extremely large resource records. The requester would
likely then allocate enough memory to store the intermediate state until reassembly is
possible, and could only detect the attack once trying to verify the signature. One potential solution to this would be to use some heuristics to determine
if a RRFRAG map makes sense. Based on what the requester could expect to receive for a query
of some form, the requester can check to see if the response it actually received fits within
those expectations. For example, if the requester indicated that it only supported Falcon-512
signatures, it can check that the advertised sizes of the fragments are no larger than 690 bytes.
We leave this issue for future exploration.

\paragraph{Unreliable networks.}
In this work we did not evaluate how \protocolname performs when UDP packets do not reach their destination.
BIND9 uses a default timeout of 800ms to determine whether it should try the request again or not, but it is unclear
if that timeout duration would make sense for \protocolname or not. This question must be answered before \protocolname
can be deployed and we leave this for future work.

\subsection{Comparing \protocolname Against Previous DNS Fragmentation Proposals}

\protocolname is not the first attempt at a
DNS-level fragmentation mechanism. Since Sivaraman's draft ``DNS message fragments" \cite{draft-muks-dns-message-fragments-00}
was not as developed as Additional Truncated Response (ATR) \cite{draft-song-atr-large-resp-03},
we will be primarily focusing on ATR in this section. ATR, Sivaraman's draft, and \protocolname,
all rely on DNS-level fragmentation. The DNS servers are required to fragment messages and
re-assemble them rather than relying on the transport layer to handle message fragmentation
for them. All three mechanisms are transport layer agnostic and could therefore be used on
both UDP and TCP. It may seem unclear why someone would want to run any of these mechanisms
over TCP, however by doing so there is the potential for sending DNS messages larger than
the 64 kilobyte maximum. ATR and Sivaraman's draft could in theory allow resource records of
~64 kilobytes to be transmitted; whereas \protocolname could allow for resource records of
arbitrary length. This is due to the difference in granularity of fragmentation that the
three mechanisms use. ATR and Sivaraman's draft fragment the DNS message as a whole, where
as \protocolname fragments individual resource records. Although there are no resource records that
require an increase to the maximum DNS message size, and therefore maximum resource record size,
it is not entirely unrealistic to see this issue potentially arising.

Before being broken \cite{EPRINT:Beullens22},
the Rainbow \cite{NISTPQC-R3:Rainbow20} post-quantum signature scheme was quite appealing due to its relatively small signature
sizes; however it had large public keys of 161600 bytes. Since DNSKEYS are sent much less frequently than
signatures, this might have been a reasonable trade off had Rainbow not been broken. It is entirely possible
that a new, secure post-quantum signature scheme is created which has similar signature and public key sizes.
(In fact, this is specifically mentioned as a desirable design characteristic in NIST's September 2022 call for additional post-quantum digital signature schemes \cite{NISTPQC-additional-signatures}.)
In order to fully support arbitrary-sized resource records, the resource record format would need to be modified to
support larger RDATA regions, and RRSIZE would need to be updated to the proper integer width.

One of the major criticisms of ATR \cite{draft-song-atr-large-resp-03} was that, since the mechanism would blindly send its additional message
as part of its response, it would cause a flood of ICMP `destination unreachable' packets to be created by
resolvers which did not support ATR. Many implementations close their sockets immediately after receiving a response,
so by the time the additional message is received the socket would no longer be accessible. This would make
debugging considerably more challenging and reduce the usefulness of ICMP messages as a whole. Another issue
arises with firewalls that have the policy of only receiving a single DNS message per query, and thus compounding
the ICMP flood issue.  \protocolname does not suffer from these issues.
Firstly, responses are only sent when they are explicitly queried for. A DNS server implementing \protocolname
will never send an additional response blindly and will never send additional messages to resolvers
that don't support \protocolname as they will never ask for them. Similarly, all DNS messages containing
RRFrags will have an associated query and will therefore not get dropped by firewalls implementing the above policy.
As \protocolname does not suffer from either of those issues, there will not be a flood of ICMP packets that
caused so much concern.

ATR also requires a slight delay between the first message being sent and the trailing messages being sent in order
to maintain message ordering. Receiving messages out of order is not an issue for \protocolname as the requesting
server will know what to expect after receiving the first message containing the RRFRAG map of the whole DNS message.
All responses after the first one will have been explicitly asked for and are not dependent on any other responses.

Where as ATR is quite lightweight, \protocolname does have some additional transportation costs. ATR costs a single
round trip plus the delay required to maintain message ordering, whereas \protocolname has $\left\lceil \frac{\text{Original response size}}{\text{Maximum message size}} \right\rceil$
round trips. With the exception of the initial round trip, these round trips can be performed in parallel, thus reducing the overall resolution time. \protocolname also requires more data to be sent, specifically as part of requesting the additional fragments.
RRFRAGs in requests are 15 bytes in size, and the number sent depends on the number of resource records, how large they are,
and how much data can fit in the maximum message size.

Sivaraman's draft \cite{draft-muks-dns-message-fragments-00} was built off of EDNS(0)'s OPT resource record requiring three fragmentation related options support to be
assigned by ICANN. \protocolname does not use the OPT pseudo-resource record and therefore does not require any options to
be defined by ICANN.

Finally both \protocolname and ATR can be implemented as a daemon on the resolver side without any changes required to the DNS software being used.
This would make deployment much simpler as it would not require a DNS operator to update their resolver software and potentially have version incompatibilities. The reassembly
could be performed entirely transparently to the resolver.

\section{Future Work}
Although \protocolname appears to be a viable solution to solving DNS message fragmentation and therefore opening the door for post-quantum DNSSEC, additional work
needs to be done. The backwards compatibility of \protocolname needs to be further explored and evaluated in real-world deployments, exploring if there are middle boxes which
cause \protocolname to fail. \protocolname as specified in this work is susceptible to memory exhaustion attacks and additional work needs to be done to prevent these attacks.
It also also likely that operators will want to limit the number of concurrent requests when using parallel \protocolname and therefore research into selecting a reasonable
limit must be done. 

In this work we provide a proof of concept daemon which transparently implements \protocolname.
Directly integrating \protocolname into DNS implementations may uncover unexpected surprises.

Our experiments only considered the case of lossless packet delivery.
In reality, UDP packet delivery is not guaranteed, so research is needed on how \protocolname behaves in unreliable networks. 
Work also needs to be done to measure any additional processing/memory overhead introduced
by \protocolname and whether that overhead is reasonable.

Any future standardization of \protocolname would depend both on \protocolname itself being evaluated by the Internet Engineering Task Force as well as appropriate post-quantum algorithms being specified for use in DNSSEC.

\section{Conclusion}

Post-quantum cryptography will inevitably need to be integrated into the DNSSEC ecosystem, however it looks like it will not
be as smooth of a transition as we would like. Of our current options, Falcon-512 is by far the most performant but even with
parallel \protocolname is still significantly slower than currently used classical signing algorithms. 
There has been recent work on shrinking Falcon-512 signatures significantly which would improve its performance. Dilithium2 is perhaps viable
as an alternative option, but considering the DNSSEC community's previous stance of ``we can avoid sending large message by shaping their contents better (smaller signatures, less additional data)'' \cite{smallersignautres}, Dilithium2 may 
receive significant resistance if proposed for use in DNSSEC. SPHINCS+-SHA256-128S is by far the worst performing of the three NIST post-quantum selections due to its slow verification and extremely large signatures which causes very large resolution times.

Message sizes are not the only thing to consider when discussing which post-quantum signing algorithm to standardize for DNSSEC, as
the security of the algorithms must also be considered. So far major attacks have been found against several candidates fairly late
in the NIST selection process. To make matters worse, those algorithms were broken with traditional computers, therefore making the attacks much more practical. Although the three selected algorithms are believed to be secure now, will they hold up to additional
scrutiny? Only time will tell. It is likely that using a hybrid of a classical signing scheme and post-quantum scheme will be desirable for some time
to ensure that the signatures are at least as strong as what are currently standardized. This will come at a further performance cost and also increase communication sizes,
and we plan to evaluate this additional cost in the future.

A final option is to wait for new post-quantum signature schemes to be invented and hope that signature sizes become more reasonable. NIST
has requested additional post-quantum signature schemes be submitted for consideration standardization \cite{NISTPQC-additional-signatures}. However, waiting several years for a better
scheme to emerge is eating into the valuable time needed to prepare for securing DNS against a quantum adversary. It is best that we plan for the
worst case of signatures sizes not improving, and be pleasantly surprised if such a scheme arises. With that in mind, we recommend
Falcon-512 as a suitable signature algorithm for use in DNSSEC with \protocolname as its delivery mechanism
to achieve reasonable resolution times.

\section*{Acknowledgments}

We gratefully acknowledge helpful discussion with Roland van Rijswijk-Deij, Andrew Fregly and Burt Kaliski, Sofía Celi, and Michael Baentsch.
D.S. was supported by Natural Sciences and Engineering Research Council of Canada (NSERC) Discovery grants RGPIN-2016-05146 and RGPIN-2022-0318, and a donation from VeriSign, Inc.

\section*{Availability}

The software implementing the daemon and experiment is available at \url{https://github.com/Martyrshot/ARRF-experiments/}.


\appendix
\section{Appendix -- Performance graphs}

\begin{figure*}[t]
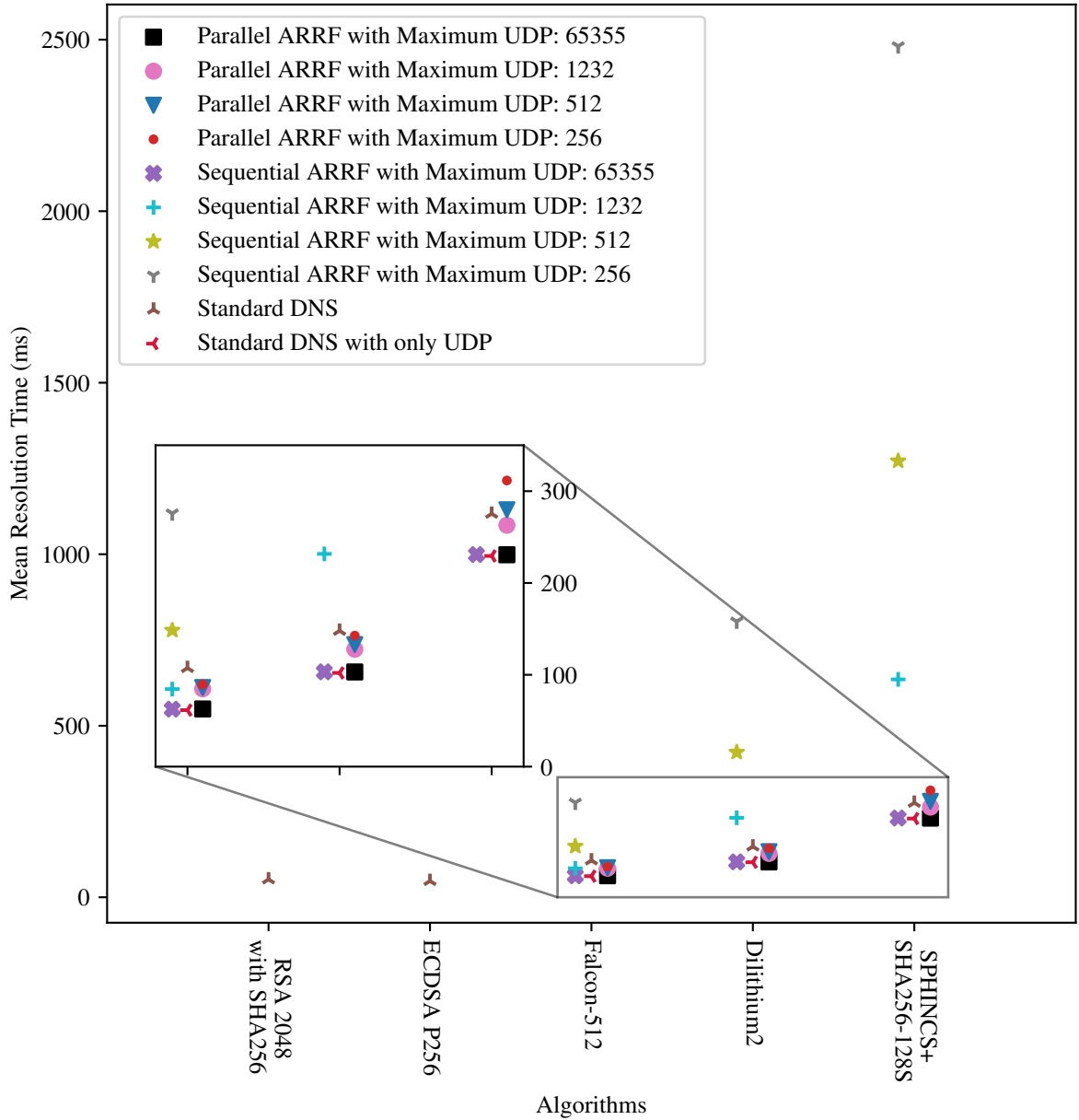

    \begin{center}
	    \begingroup\makeatletter
\makeatother\endgroup
    \end{center}\caption{Mean resolution times in milliseconds with 10ms latency and 128 kilobytes per second bandwidth}
	\label{fig:scatter-10ms-128kbps}
\end{figure*}

\begin{figure*}[t]
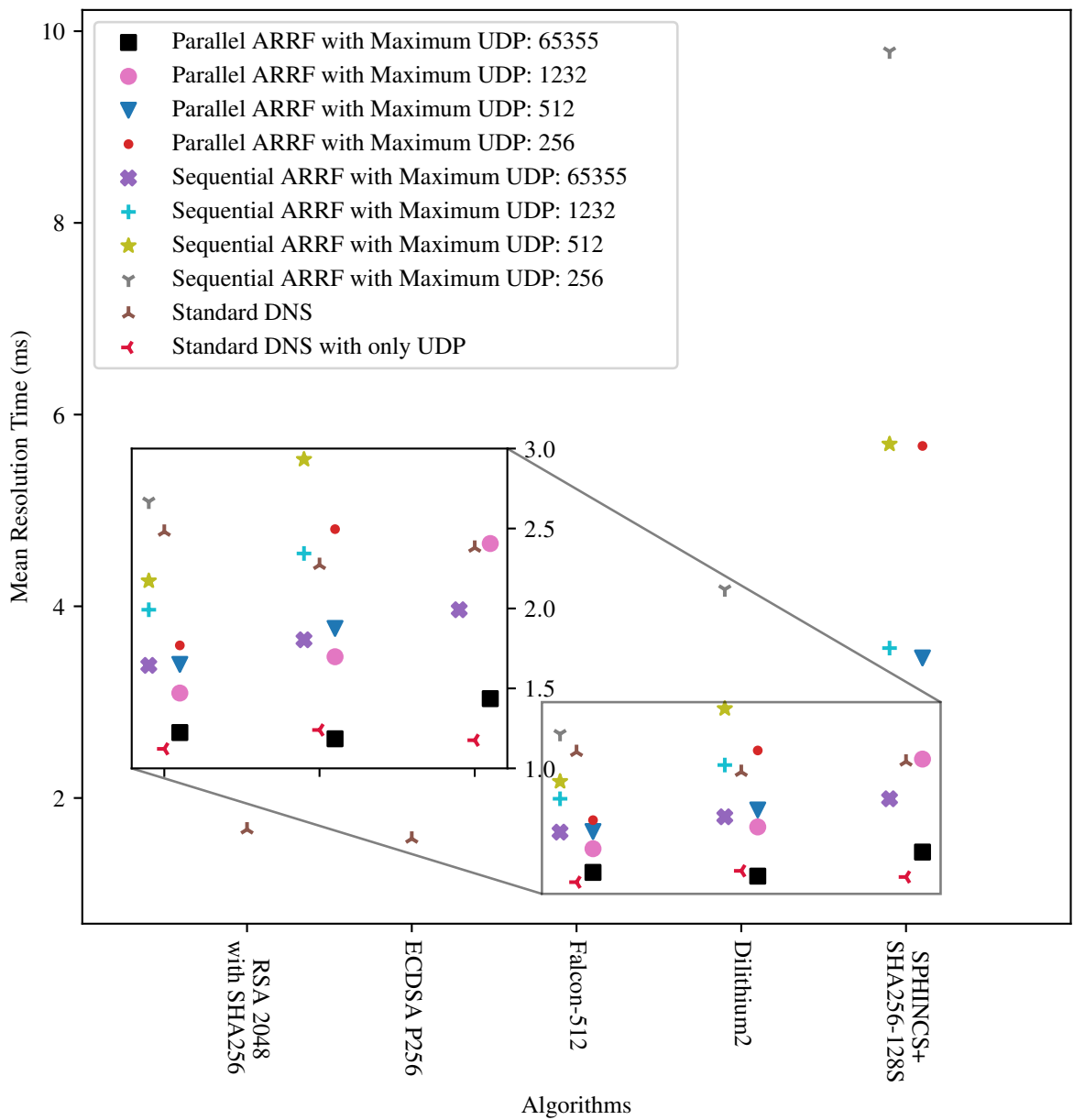

    \begin{center}
        \begingroup\makeatletter%
\makeatother\endgroup
    \end{center}
	\caption{Mean resolution times in milliseconds with 0ms latency and unlimited bandwidth}
	\label{fig:scatter-0ms-unlimited}
\end{figure*}

Figures \ref{fig:scatter-10ms-128kbps}, \ref{fig:scatter-10ms-50mbps}, \ref{fig:scatter-100ms-50mbps}, and \ref{fig:scatter-0ms-unlimited} visualize the performance of ARRF in batched and sequential mode in various network scenarios and at different maximum UDP packet sizes compared with standard DNS with TCP fallback or UDP only mode.

\begin{figure*}[t]
    \begin{center}
        \begingroup\makeatletter%
\makeatother\endgroup
    \end{center}
	\caption{Mean resolution times in milliseconds with 10ms latency and 50 megabytes per second bandwidth}
	\label{fig:scatter-10ms-50mbps}
\end{figure*}

\begin{figure*}[t]
    \begin{center}
        \begingroup\makeatletter%
\makeatother\endgroup
    \end{center}
	\caption{Mean resolution times in milliseconds with 10ms latency and 50 megabytes per second bandwidth}
	\label{fig:scatter-100ms-50mbps}
\end{figure*}

\end{document}